\documentstyle[psfig]{article}
\begin{document}
\title{%
Theory of the Transient and Stationary Fluctuation Theorems for
a Dragged Brownian Particle in a Fluid
}
\author{R. van Zon and E.G.D. Cohen\\The Rockefeller University\\1230 York Ave., New York NY 10021-6399}
\date{22 October 2002}
\maketitle

{\bf Comment on ``Experimental Demonstration of Violations of the }
{\bf Second Law of Thermodynamics for Small Systems and Short Time Scales''}
\cite{Wangetal02}.
{\ }\\

In a recent experimental demonstration of the {\em transient
fluctuation theorem} (TFT) by Wang {\it et al.}\cite{Wangetal02}, a
small latex bead, initially at rest in a harmonic trap and in
equilibrium with a surrounding fluid, starts to be dragged at a
velocity ${\bf v}_{opt}$ at $t=0$ by means of the trap. Measurements
of the entropy production over a time $\tau$,
\begin{equation}
	\Sigma_\tau\equiv\frac{1}{k_BT}\,{\bf v}_{opt}\cdot\int_0^\tau {\bf F}(t) \,dt ,
\label{entropyproductiondefinition} 
\end{equation}
were made, with $T$ the temperature, $k_B$ Boltzmann's constant, ${\bf
F}(t)=-k({\bf x}(t)-{\bf x}_0(t))$, ${\bf x}(t)$ the bead's position,
and ${\bf x}_0(t)$ zero for $t<0$, and ${\bf v}_{opt} t$ for
$t>0$. Thus, $P_T(\Sigma_\tau<0)$ and $P_T(\Sigma_\tau>0)$, the
probabilities to find a negative resp. positive values of
$\Sigma_\tau$, were measured and shown to comply with the
integrated TFT\cite{Wangetal02,EvansSearles94}:
\begin{equation}
	\frac{P_T(\Sigma_\tau<0)}{P_T(\Sigma_\tau>0)}=\langle
\exp[-\Sigma_\tau]\rangle_{T+},
\label{ITFT}
\end{equation}
where on the right-hand side (r.h.s.)\ the average is performed over
transient ($T$) trajectories with positive $\Sigma_\tau$.

Assuming the probability distribution $P({\bf x},t)$ is described by
the Smoluchovski equation, it satisfies
\begin{equation}
	\partial_t P({\bf x},t) 
	= \nabla\cdot\{D\nabla + \tau_r^{-1}[{\bf x}-{\bf x}_0(t)]\}
		P({\bf x},t).
\label{Smoluchovski}
\end{equation}
with $D$ is the diffusion constant and $\tau_r$ the particle's
relaxation time in the harmonic trap.  Since
Eq.~(\ref{entropyproductiondefinition}) is a linear combination of
positions at different times, and the Greens function and equilibrium
state are Gaussian, so will the $\Sigma_\tau$ distribution.  It can
then be characterized by its first two moments
\begin{eqnarray}
	\langle \Sigma_\tau\rangle_T
		&=& \sigma\tau-\sigma\tau_r(1-e^{-\tau/\tau_r})
\nonumber
\\
	\langle \Delta\Sigma_\tau^2\rangle_T
	&=&
		\left\langle [ \Sigma_\tau -
		\langle\Sigma_\tau\rangle
		]^2
		\right\rangle_T
		= 2 \langle \Sigma_\tau\rangle_T.
\label{4}
\end{eqnarray}
Here $\sigma=|{\bf v}_{opt}|^2/D$ is the entropy production rate in
the non-equilibrium stationary state (NESS).  Eq.~(\ref{ITFT}) follows
from Eq.~(\ref{4}) and the Gaussian distribution of
$\Sigma_\tau$. Sevick {\it et al.\/} have independently developed an
equivalent theory for the TFT from a Langevin approach\cite{note}.

Although the TFT looks similar to the {\em stationary state
fluctuation theorem}\cite{GallavottiCohen95} (SFT), it is quite
different\cite{CohenGallavotti99}.  The TFT looks at variations in
entropy production over a trajectory of length $\tau$ starting from an
initial equilibrium ensemble.  The SFT deals with the distribution of
entropy production $\Sigma_\tau$ over trajectory segments of length
$\tau$ along a single trajectory in the NESS.  As above, the SFT can be
derived from Eq.~(\ref{Smoluchovski}) because the NESS is Gaussian, as
is the stationary ($S$) distribution of $\Sigma_\tau$. One obtains
\begin{equation}
	\langle\Sigma_\tau\rangle_S =  \sigma \tau ;\quad 
	\langle\Delta\Sigma^2_\tau\rangle_S 
	=
	\langle\Delta\Sigma^2_\tau\rangle_T 
\label{5}
\end{equation}

\begin{figure}[!t]
\centerline{\psfig{file=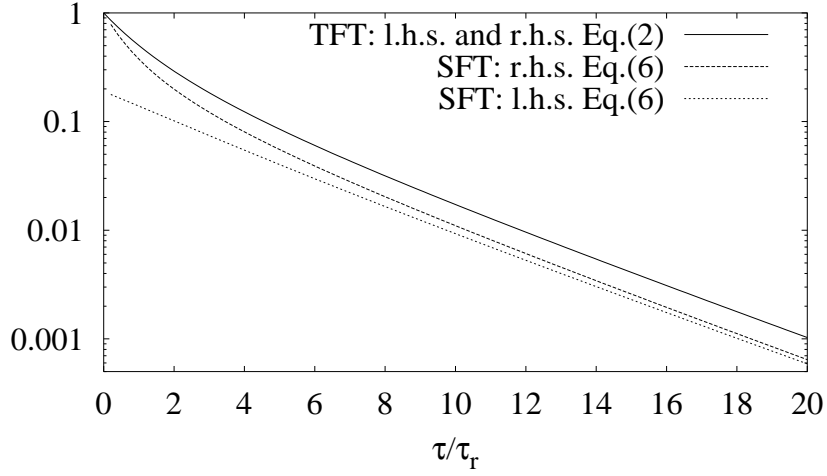,width=\textwidth}}
%\centerline{\raisebox{14pt}{$\tau$}}
\caption{l.h.s.\ and r.h.s.\ of Eq.~(\ref{ITFT}) (solid line) and
Eq.~(\ref{ISFT}) (dotted line and dashed line, resp.), for $\sigma\tau_r=1$.}
\label{figure}
\end{figure}

\noindent
Now  $\langle\Delta\Sigma^2_\tau\rangle_S\rightarrow
2\langle\Sigma_\tau\rangle_S$ only as $\tau\rightarrow\infty$
[Eqs.~(\ref{4},\ref{5})],
so
\begin{eqnarray}
	\frac{P_S(\Sigma_\tau<0)}{P_S(\Sigma_\tau>0)}
&\stackrel{\tau\rightarrow\infty}{=}&
\langle \exp[-\Sigma_\tau]\rangle_{S+},
\label{ISFT}
\end{eqnarray}

Fig.~\ref{figure} shows the left-hand side (l.h.s.)\ and r.h.s.\ of
Eqs.~(\ref{ITFT}), and (\ref{ISFT}); for $\tau\rightarrow\infty$, the
curves of the SFT indeed approach each other[Eq.~(\ref{ISFT})], but
the curves of the TFT and of the SFT approach a constant, non-tivial
ratio
\begin{eqnarray*}
	\frac{P_T(\Sigma_\tau<0)}{P_T(\Sigma_\tau>0)}\left/
	\frac{P_S(\Sigma_\tau<0)}{P_S(\Sigma_\tau>0)}\right.
\stackrel{\tau\rightarrow\infty}{=}
	\exp\left[\frac{\sigma\tau_r}{2}\right],
\end{eqnarray*}
as follows from the asymptotics of the error functions in the explicit
expressions for the left-hand sides of Eqs.~(\ref{ITFT}) and
(\ref{ISFT}).  So a negative 
%entropy production 
$\Sigma_\tau$
is less likely in the
NESS than in the transient state.  In fact for large $\sigma\tau_r$,
negative fluctuations in the NESS are exponentially suppressed.  This
is important if one wants to devise an experiment on the SFT.
More details will be given in a future publication.

This work has been supported by the
U.S. Department of Energy, under grant No. DE-FG-02-88-ER13847.

%{\ }\\
%\noindent
%R.~van~Zon and E.G.D.~Cohen
%
%	The Rockefeller University
%
%	1230 York Avenue
%
%	New York, NY 10021-6399

\end{document}